\documentclass[11pt]{article}
\usepackage{graphicx}

\newcommand{\BABARPubYear}    {05}

\newcommand{\BABARConfNumber} {013}
\newcommand{\SLACPubNumber} {11433}
\input pubboard/babarsym

\def\pipi     {\ensuremath{\pi\pi}}

\def\akpi  {\ensuremath{{\cal A}_{K\pi}}}

\def\spipi {\ensuremath{S_{\pi\pi}}}
\def\cpipi {\ensuremath{C_{\pi\pi}}}
\def\de {\ensuremath{\Delta E}}

\long\def\inst#1{\par\nobreak\kern 4pt\nobreak
    {\it #1}\par\vskip 10pt plus 3pt minus 3pt}

\begin{document}
{\pagestyle{empty}

\begin{flushright}
\babar-CONF-\BABARPubYear/\BABARConfNumber \\
SLAC-PUB-\SLACPubNumber \\
%hep-ex/\LANLNumber \\
%\today \\
\end{flushright}

\par\vskip 5cm

% Title of the paper
\begin{center}
\Large \bf \boldmath Improved Measurements of Branching Fractions for\\
 $\Bz\to\pip\pim$, $\Kp\pim$, and Search for $\Kp\Km$
at \babar\
\end{center}
\bigskip

\begin{center}
\large The \babar\ Collaboration\\
\mbox{ }\\
\today
\end{center}
\bigskip \bigskip

% Abstract
\begin{center}
\large \bf Abstract
\end{center}
We present preliminary measurements of branching fractions 
for the charmless two-body decays $\Bz\to\pip\pim$ and $\Kp\pim$, and
a search for $\Bz\to\Kp\Km$ using a data sample of approximately 227 
million $\BB$ decays.  Signal yields are extracted with 
a multi-dimensional maximum likelihood fit, and the efficiency is
corrected for the effects of final-state radiation.  We find the
charge-averaged branching fractions (in units of $10^{-6}$):
\begin{eqnarray}
\BR(\Bz\to\pip\pim) & = & 5.5\pm 0.4\pm 0.3, \\
\BR(\Bz\to\Kp\pim)  & = & 19.2\pm 0.6\pm 0.6, \\
\BR(\Bz\to\Kp\Km)   & = & < 0.40.
\end{eqnarray}
The errors are statistical followed by systematic, and the upper limit on
$\Kp\Km$ represents a confidence level of $90\%$.

\vfill
\begin{center}
Presented at the 
International Europhysics Conference On High-Energy Physics (HEP 2005),
7/21---7/27/2005, Lisbon, Portugal
\end{center}

\vspace{1.0cm}
\begin{center}
{\em Stanford Linear Accelerator Center, Stanford University, 
Stanford, CA 94309} \\ \vspace{0.1cm}\hrule\vspace{0.1cm}
Work supported in part by Department of Energy contract DE-AC03-76SF00515.
\end{center}

\newpage
} % end of pagestyle{empty}

% Input author list file
\begin{center}
\small

The \babar\ Collaboration,
\bigskip

B.~Aubert,
R.~Barate,
D.~Boutigny,
F.~Couderc,
Y.~Karyotakis,
J.~P.~Lees,
V.~Poireau,
V.~Tisserand,
A.~Zghiche
\inst{Laboratoire de Physique des Particules, F-74941 Annecy-le-Vieux, France }
E.~Grauges
\inst{IFAE, Universitat Autonoma de Barcelona, E-08193 Bellaterra, Barcelona, Spain }
% ...

\end{center}\newpage

% The body of the paper starts here
\section{INTRODUCTION}
\label{sec:Introduction}
Charmless hadronic two-body $B$ decays to pions and kaons provide a wealth of 
information on \CP\ violation in the $B$ system, including all three angles of the 
unitarity triangle.  The time-dependent \CP\ asymmetries in the $\pi\pi$ system can be 
used to measure the angle $\alpha$~\cite{pipialpha}; the decay rates for the 
$K\pi$ channels provide information on $\gamma$~\cite{gammaKpi}; and the 
time-dependent \CP\ asymmetry in $\piz\KS$ approximately measures $\beta$ in the 
standard model~\cite{kspi0} and is a sensitive probe of new physics in the $b\to s$ 
penguin-decay process~\cite{kspi0NP}.  Recently, direct \CP\ violation in decay was 
established in the $B$ system through observation of a significant rate asymmetry between 
$\Bz\to\Kp\pim$ and $\Bzb\to\Km\pip$~\cite{babarAkpi,belleAkpi}.  
As $B$-physics experiments accumulate much larger data sets, charmless two-body $B$ decays 
will continue to play a critical role in testing the standard model description of 
\CP\ violation.

In order to extract the maximum information from these decays it is necessary to
understand the underlying hadron dynamics, and measurements of branching fractions for 
all of the charmless two-body $B$ decays involving combinations of $\pipm,~\Kpm,~\piz$, 
and $\KS$ are invaluable in testing the various theoretical 
approaches~\cite{thy}.  We present preliminary measurements of branching fractions 
for the decays~\cite{cc} $\Bz\to\pip\pim$ and $\Kp\pim$, and a search for the decay $\Bz\to\Kp\Km$
using a data set $2.5$ times larger than the one used for our previous measurements of
these quantities~\cite{BaBarsin2alpha2002}.
Table~\ref{tab:oldresults} summarizes previous experimental 
measurements~\cite{BaBarsin2alpha2002,belleBR,cleoBR} and current theoretical 
estimates of the branching fractions for these decays.

\begin{table}[!b]
\caption{Summary of existing branching fraction measurements (in units of $10^{-6}$) and 
theoretical estimates for the decays $\Bz\to\pip\pim,~\Kp\pim,~\Kp\Km$.  Theory estimates 
are from Beneke {\em et al.} and Keum in Ref.~\cite{thy}.}
\begin{center}
\begin{tabular}{ccccc}
\hline\hline
   Mode    & $\BR({\rm \babar})$~\cite{BaBarsin2alpha2002} & $\BR({\rm Belle})$~\cite{belleBR} & $\BR({\rm CLEO})$~\cite{cleoBR} & Theory\\\hline
$\pip\pim$ & $4.7\pm 0.6\pm 0.2$ & $4.4\pm 0.6\pm 0.3$ & $4.5^{+1.4~+0.5}_{-1.2~-0.4}$ & $4.6$-$11.0$\\
$\Kp\pim$  & $17.9\pm 0.9\pm 0.7$ & $18.5\pm 1.0\pm 0.7$ & $18.0^{+2.3~-1.2}_{-2.1~-0.9}$ & $12.7$-$21.0$\\
$\Kp\Km$   & $<0.6$ & $<0.7$ & $<0.8$ & $0.007$-$0.080$ \\
\hline\hline
\end{tabular}
\label{tab:oldresults}
\end{center}
\end{table}

\section{THE \babar\ DETECTOR AND DATASET}
\label{sec:babar}
The data sample used for this search contains $(226.6\pm 2.5)\times 10^6$ 
$\Y4S\to\BB$ decays collected by the \babar\ detector~\cite{ref:babar} at the 
SLAC PEP-II $\epem$ asymmetric-energy storage ring.  The primary detector 
components used in the analysis are a charged-particle tracking system 
consisting of a five-layer silicon vertex detector and a 40-layer drift chamber 
surrounded by a $1.5$-T solenoidal magnet, and a dedicated particle-identification 
system consisting of a detector of internally reflected 
Cherenkov light (DIRC) providing $K$--$\pi$ separation over the range of laboratory 
momentum relevant for this analysis ($1.5$--$4.5\gevc$).

\section{ANALYSIS METHOD}
\label{sec:Analysis}
The data sample used in this analysis is similar to that used in the \babar\ measurements 
of direct \CP\ violation in $\Kp\pim$~\cite{babarAkpi} and time-dependent
\CP-violating asymmetry amplitudes $\spipi$ and $\cpipi$~\cite{BaBarsin2alpha2005} (the
reader is referred to those references for further details on the analysis
technique).  Relative to the event selection applied in the \CP\ analyses, we remove 
the requirement on the difference in the decay times ($\Delta t$) between the two $B$ 
mesons in order to minimize systematic uncertainty on the branching fraction measurements.  
All other selection criteria are identical to those applied in 
Refs.~\cite{babarAkpi,BaBarsin2alpha2005}.  We identify $B\to h^+h^-$ ($h = \pi$ or $K$) candidates 
with selection requirements on track and Cherenkov-angle ($\theta_c$) quality, $B$-decay 
kinematics, and event topology, and determine signal and background yields through a 
multi-dimensional maximum-likelihood fit.  The final sample contains $69264$ events and 
is defined by requirements on the energy difference, $\left | \de\right | < 150\mev$, and 
energy-substituted mass, $5.20 < \mes < 5.29\gevcc$, of the selected $B$ 
candidates~\cite{ref:mesdedef}.

The efficiency of the selection criteria is determined in large samples of 
\geant-based Monte Carlo simulated signal decays.  We include the
effects of electromagnetic radiation from charged particles using the PHOTOS
simulation package~\cite{photos}.  The addition of final-state radiation (FSR) leads
to the development of a low-energy tail in the distribution of $\de$ for $\Bz\to h^+h^-$
signal candidates, which can cause some fraction of events to fail the 
$\left |\de \right | < 150\mev$ requirement.  We have implemented a detailed QED 
calculation~\cite{QED} up to ${\cal O}(\alpha)$ in order to correct the efficiency obtained by the 
PHOTOS simulation.  Table~\ref{tab:eff} summarizes the comparison of the efficiencies for 
the different modes assuming no FSR, the PHOTOS result, and the QED calculation.  For the 
branching fraction measurement we use the efficiency as determined by the QED calculation, 
and take the difference with respect to PHOTOS as the systematic uncertainty.

\begin{table}[!tbp]
\caption{Summary of total detection efficiencies $(\%)$ for signal decays determined in
\geant\ Monte Carlo samples without FSR effects, compared with the results using
PHOTOS and the leading-order QED calculation.  We use the latter result in calculating
the branching fraction and take the difference with PHOTOS as the systematic 
uncertainty.  Uncertainties are statistical only.}
\begin{center}
\begin{tabular}{cccc}
\hline\hline
   Mode    & No FSR & PHOTOS & QED\\\hline
$\pip\pim$ & $40.9\pm 0.2$ & $39.9\pm 0.2$ & $39.4\pm 0.2$\\
$\Kp\pim$  & $39.9\pm 0.2$ & $38.9\pm 0.2$ & $38.4\pm 0.2$\\
$\Kp\Km$   & $38.6\pm 0.3$ & $37.8\pm 0.3$ & $37.6\pm 0.3$\\
\hline\hline
\end{tabular}
\label{tab:eff}
\end{center}
\end{table}

In addition to signal $\pip\pim$, $\Kp\pim$, and (possibly) $\Kp\Km$ events, the selected
sample includes background from the process $\epem\to q\bar{q}~(q = u,d,s,c)$.  Possible backgrounds
from other $B$ decays are small relative to the signal yields ($<1\%$), and are treated as a 
systematic uncertainty.  We use an unbinned, extended maximum-likelihood fit to extract 
simultaneously signal and background yields in the three topologies ($\pi\pi$, $K\pi$, and $KK$).  
The fit uses the discriminating variables $\mes$, $\de$, $\theta_c$, and the Fisher discriminant 
${\cal F}$ described in Ref.~\cite{BaBarsin2alpha2002}, where the likelihood for event $j$ is 
obtained by summing the product of the event yield $n_i$ and probability ${\cal P}_i$ over the 
signal and background hypotheses $i$.  The total likelihood for the sample is
\begin{equation}
{\cal L} = \exp{\left(-\sum_{i}n_i\right)}
\prod_{j}\left[\sum_{i}n_i{\cal P}_{i}(\vec{x}_j;\vec{\alpha}_i)\right].
\end{equation}
The probabilities ${\cal P}_i$ are evaluated as the product of 
the probability density functions (PDFs) with parameters $\vec{\alpha_i}$,
for each of the independent variables 
$\vec{x}_j = \left\{\mes, \de, {\cal F}, \theta_c^+,\theta_c^-\right\}$,
where $\theta_c^+$ and $\theta_c^-$ are the Cherenkov angles for the 
positively- and negatively-charged tracks, respectively.  The largest correlation
between the $\vec{x}_j$ is $13\%$ for the pair $(\mes,\de)$ and we have confirmed
that it has negligible effect on the fitted yields.  For both signal and background, 
the ${\Kpm\pimp}$ yields are parameterized as 
$n_{\Kpm\pimp} = n_{K\pi}\left(1\mp\akpi\right)/2$, and we fit directly
for the total yield $n_{K\pi}$ and the asymmetry $\akpi$.  The result for
$\akpi$ is used only as a consistency check and does not supersede our previously
published result~\cite{babarAkpi}.

The eight parameters describing the background shapes for $\mes$, $\de$, and ${\cal F}$ are all
allowed to vary freely in the maximum-likelihood fit.  We use a threshold function~\cite{argus}
for $\mes$ ($1$ parameter),  a second-order polynomial for $\de$ ($2$ parameters), and
a sum of two Gaussian distributions for ${\cal F}$ ($5$ parameters).  For the signal shape in 
$\mes$, we use a single Gaussian distribution to describe all three channels and allow the mean 
and width to vary freely in the fit.  For $\de$, we use the sum of two Gaussian distributions
(core $+$ tail), where the core parameters are common to all channels and allowed to vary freely, 
and the tail parameters are determined separately for each channel from Monte Carlo simulation 
and fixed in the fit.  Given that the tail is dominated by FSR effects, we take the shape directly 
from the Monte Carlo samples after correcting for the difference between PHOTOS and the QED 
calculation.  For the signal shape in ${\cal F}$, we use an asymmetric Gaussian function with 
different widths below and above the mean.  All three parameters are determined in Monte Carlo 
simulation and fixed in the maximum-likelihood fit.  The $\theta_c$ PDFs are obtained from a 
sample of approximately $430000$ $D^{*+}\to D^0\pi^+\,(\Dz\to\Km\pip)$ decays reconstructed in 
data, where $\Kmp/\pipm$ tracks are identified through the charge correlation with the $\pipm$ 
from the $D^{*\pm}$ decay.  The PDFs are constructed separately for $\Kp$, $\Km$, $\pip$, and 
$\pim$ tracks as a function of momentum and polar angle using the measured and expected values of 
$\theta_c$, and its uncertainty.  We use the same PDFs for signal and background events.

Table~\ref{tab:results} summarizes the fitted signal and background yields, and $K\pi$ 
charge asymmetries.  We find a value of $\akpi$ consistent with our previously published result, 
and a background asymmetry consistent with zero.  The signal yields are somewhat higher than the 
values reported in Ref.~\cite{babarAkpi} due to the removal of the $\Delta t$ selection 
requirement and the addition of the radiative tail in the signal $\de$ PDF.  In order to quantify 
the effect of FSR on the fitted yields, we perform a second fit using a single Gaussian for the 
$\de$ PDF allowing the mean and width to vary freely.  The results are shown in the second column 
of Table~\ref{tab:results}, where we find that ignoring FSR lowers the $\pi\pi$ yield by $4.5\%$ 
and the $K\pi$ yield by $2.4\%$.

\begin{table}[!tbp]
\caption{Summary of the branching fraction fit using a sample of approximately $227$ million $\BB$ pairs.
For comparison, we show the results using a single Gaussian for the signal $\de$ PDF, which would correspond
to an analysis that ignores FSR effects.}
\begin{center}
\begin{tabular}[!tbp]{ccc}
\hline
\hline
 Parameter                 &	Nominal Fit        & Ignoring FSR\\
\hline
$N_{\pi\pi}$               &$    491 \pm 35     $&$   469\pm 34       $ \\
$N_{K\pi}$                 &$   1674 \pm 53     $&$  1634\pm 52       $ \\
${\cal A}_{K\pi}$          &$  -0.135\pm 0.030  $&$  -0.135\pm 0.030  $ \\
$N_{KK}$                   &$     3.0\pm 13.1   $&$	5.3\pm 12.6   $ \\
$N_{b\pi\pi}$              &$   32977\pm 194    $&$   32998\pm 194    $ \\
$N_{bK\pi}$                &$   20761\pm 169    $&$   20801\pm 169    $ \\
${\cal A}_{bK\pi}$         &$   0.002\pm 0.008  $&$   0.002\pm 0.008  $ \\
$N_{bKK}$                  &$   13358\pm 126    $&$   13356\pm 126    $ \\
\hline\hline
\end{tabular}			
\label{tab:results}
\end{center}
\end{table}

\begin{figure}[!tbp]
\begin{center}
\includegraphics[width=0.32\linewidth]{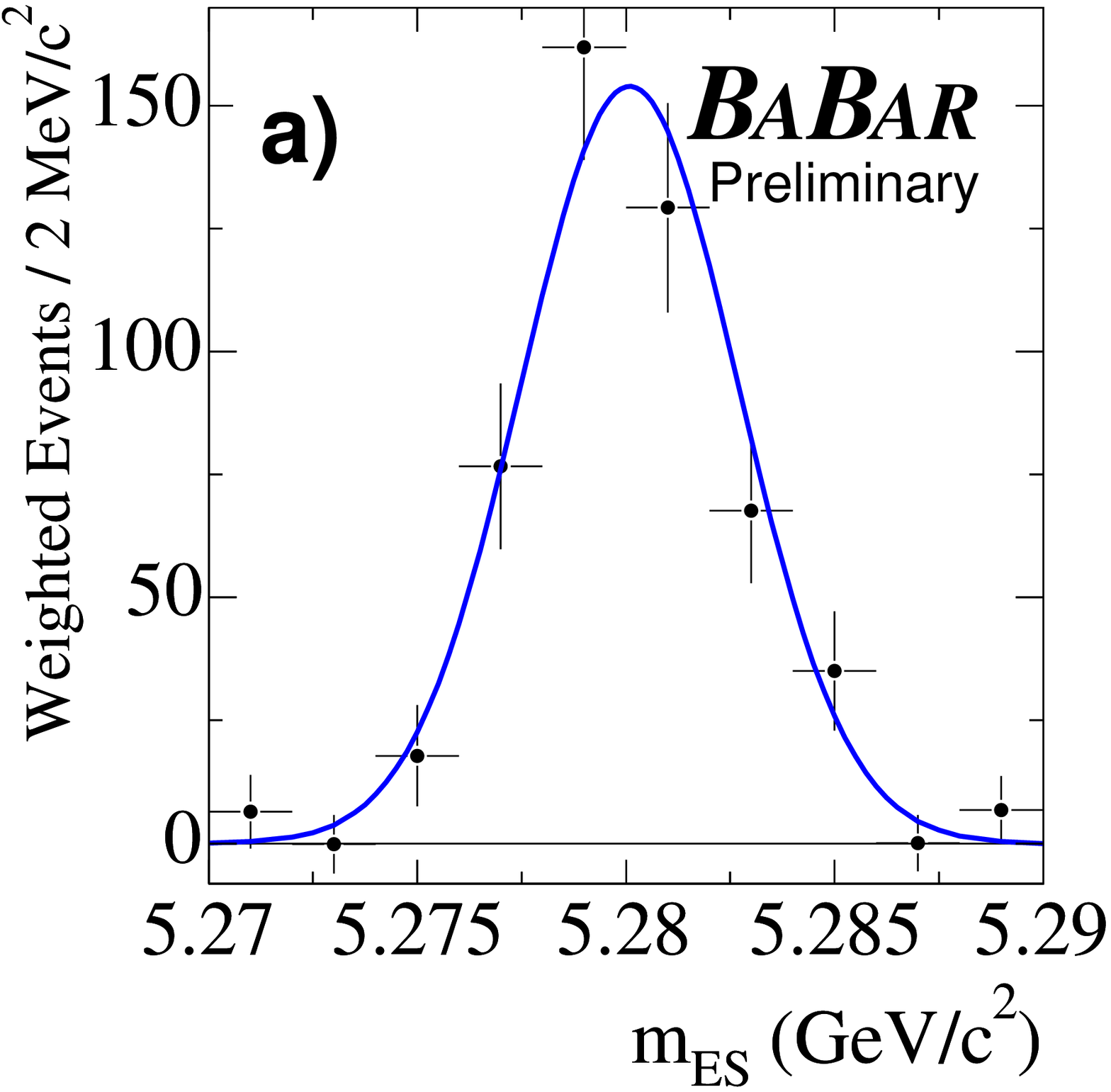}
\includegraphics[width=0.32\linewidth]{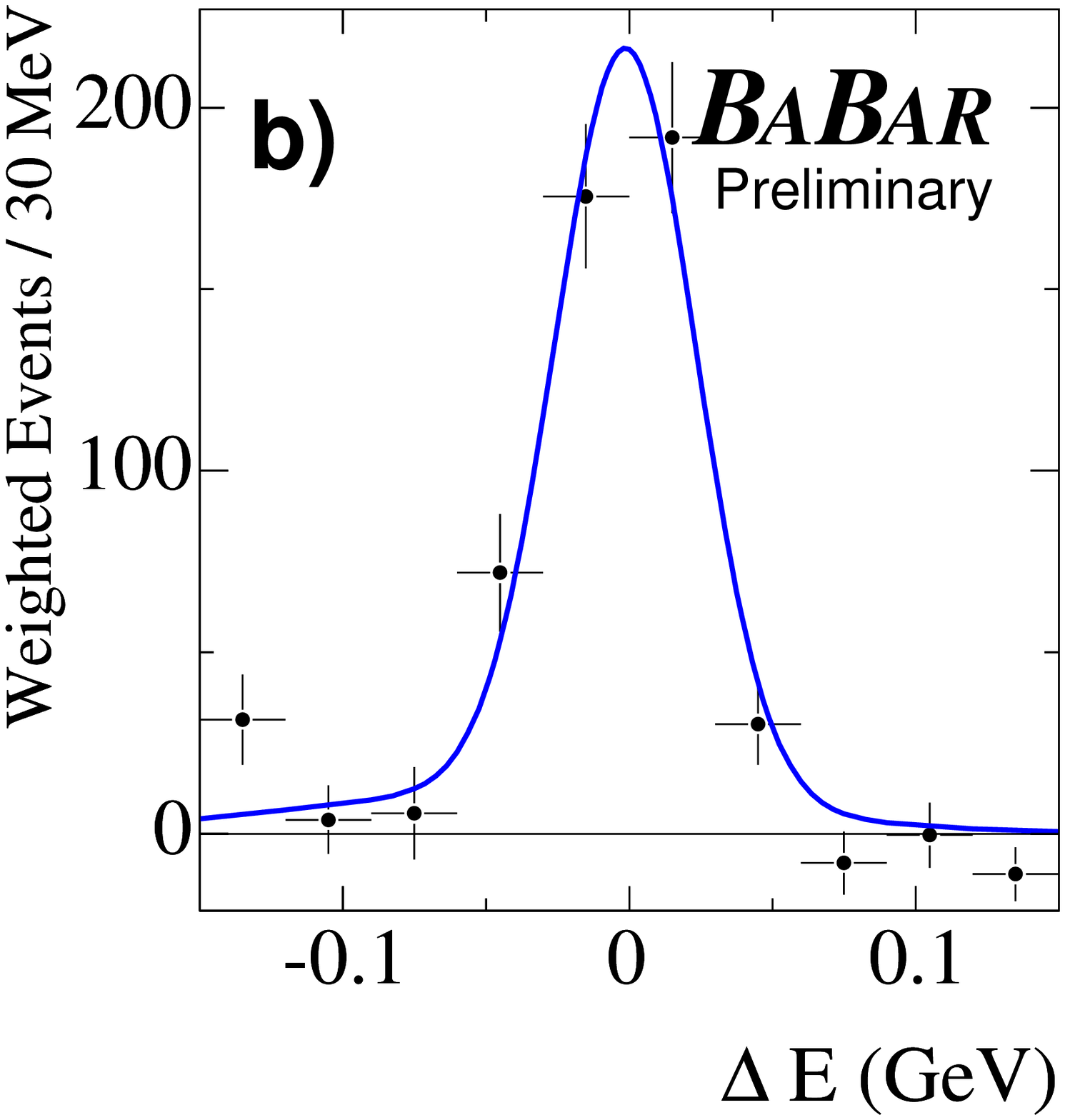}
\includegraphics[width=0.32\linewidth]{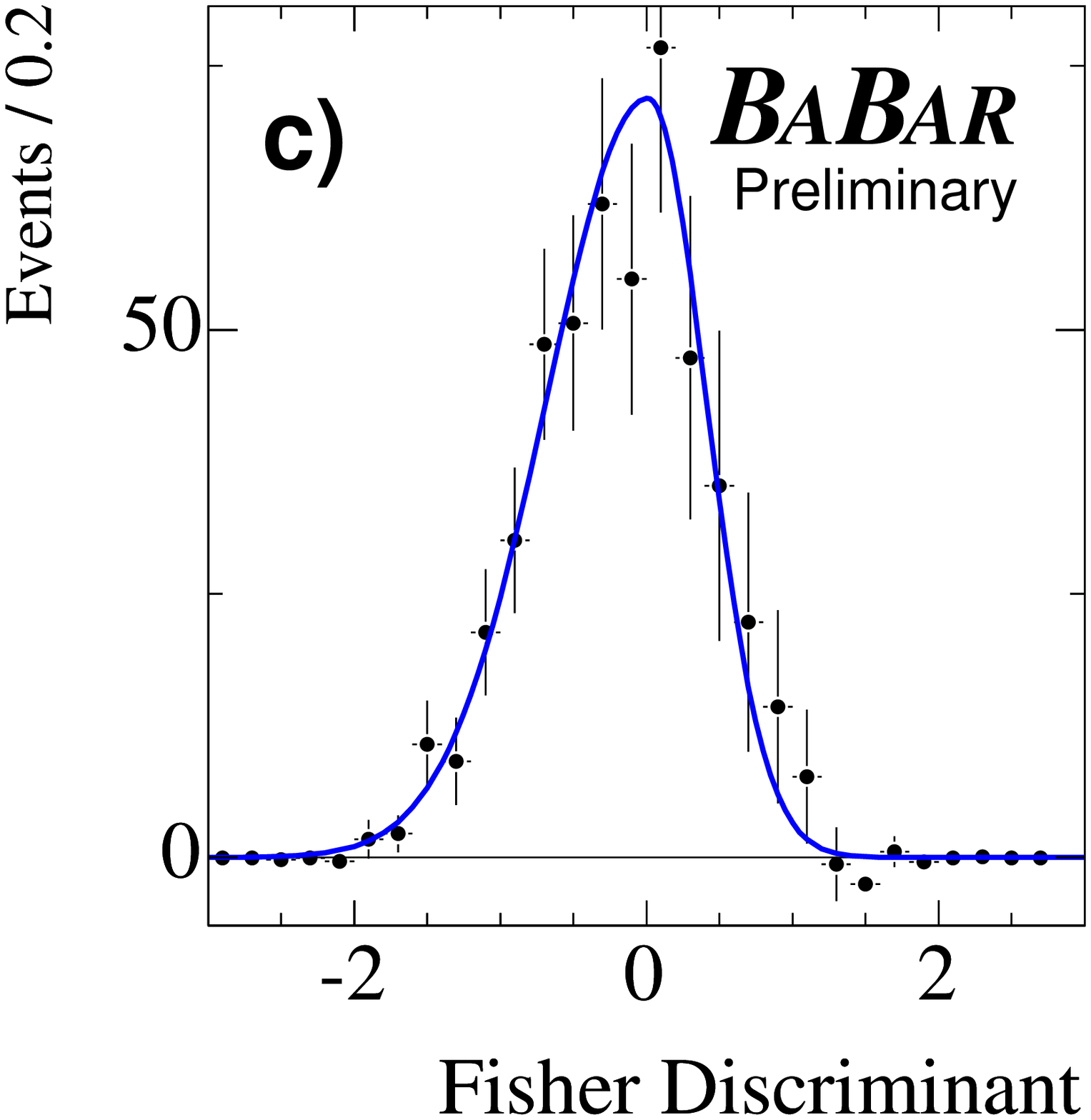}
\includegraphics[width=0.32\linewidth]{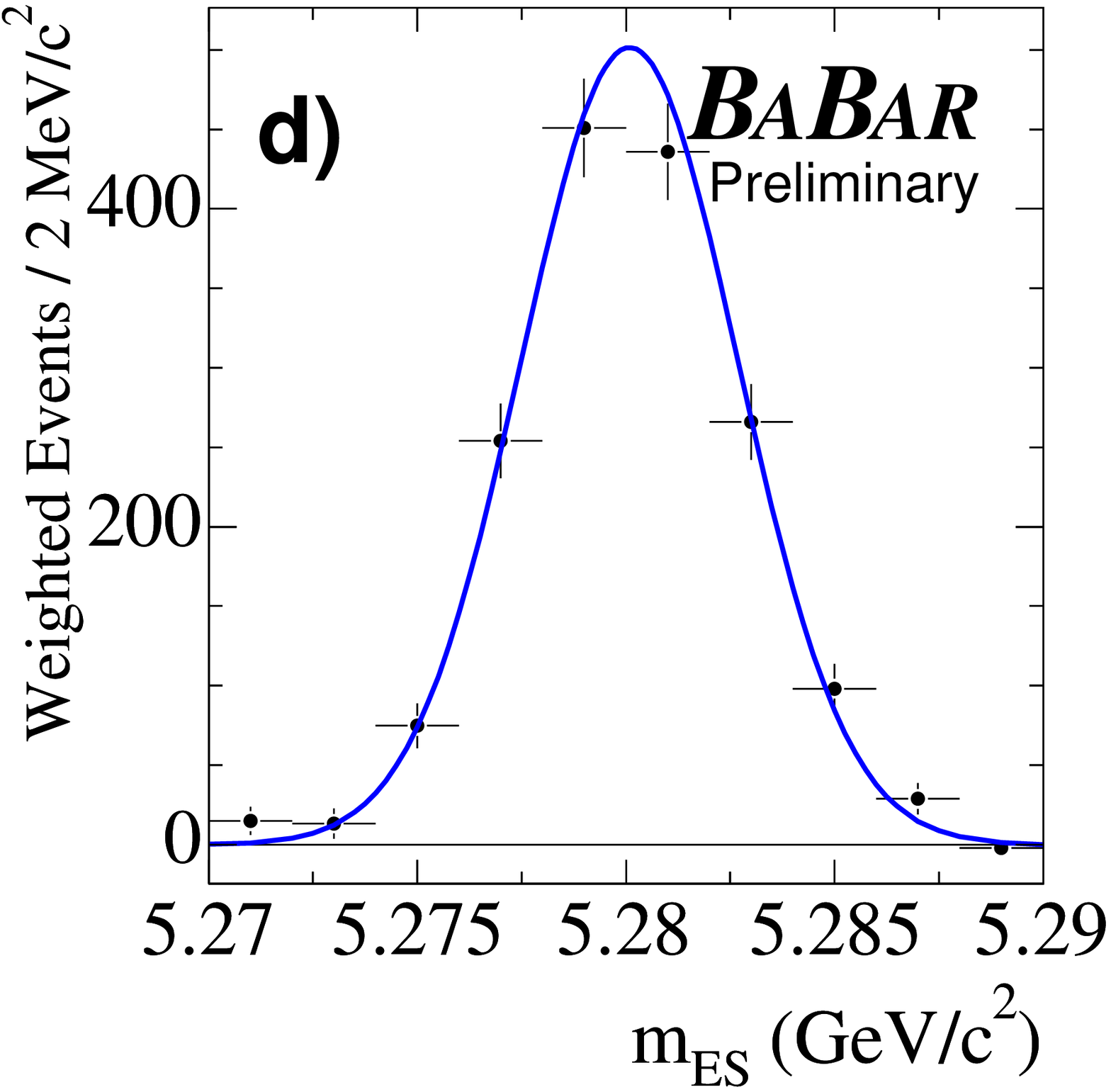}
\includegraphics[width=0.32\linewidth]{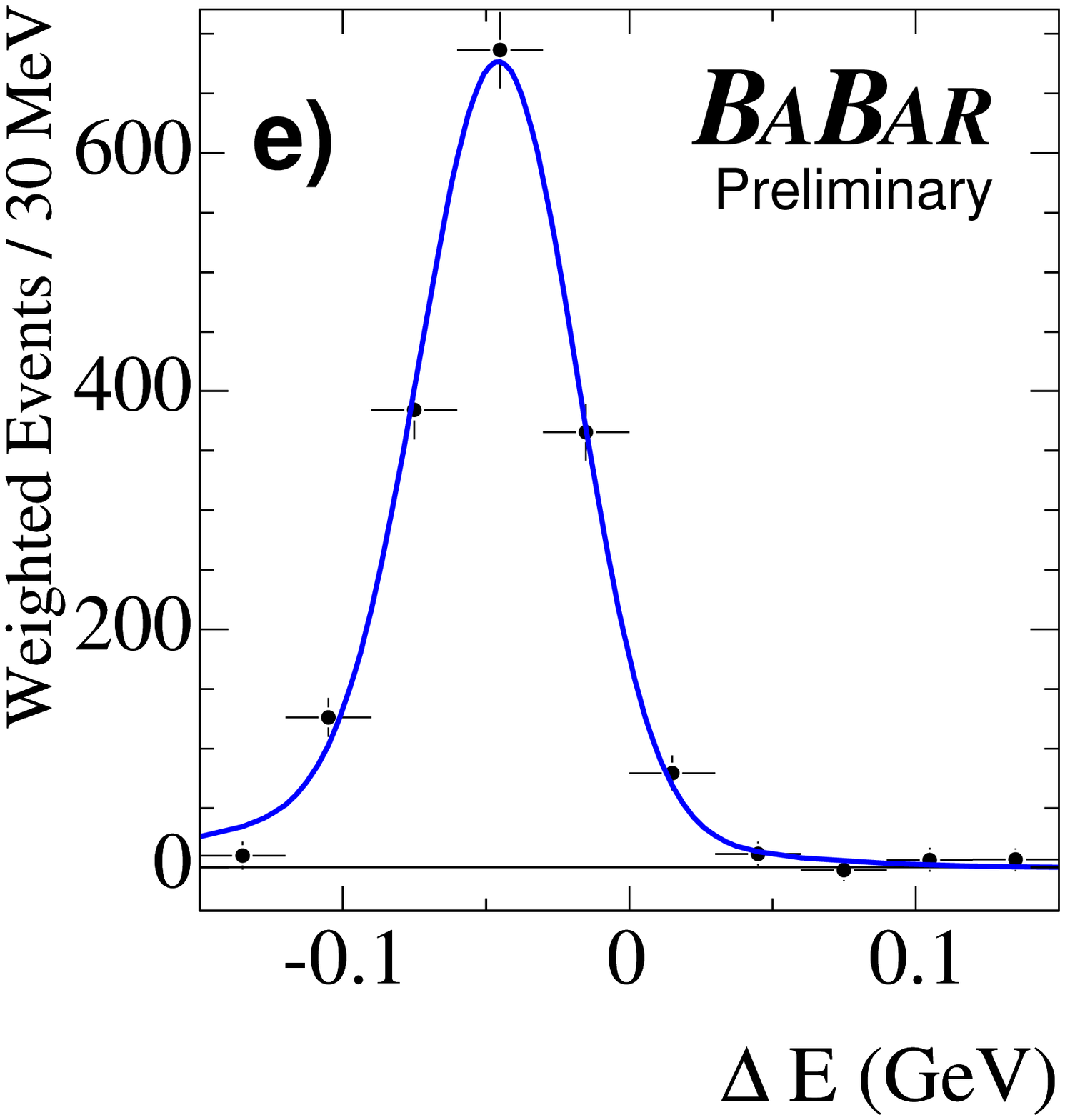}
\includegraphics[width=0.32\linewidth]{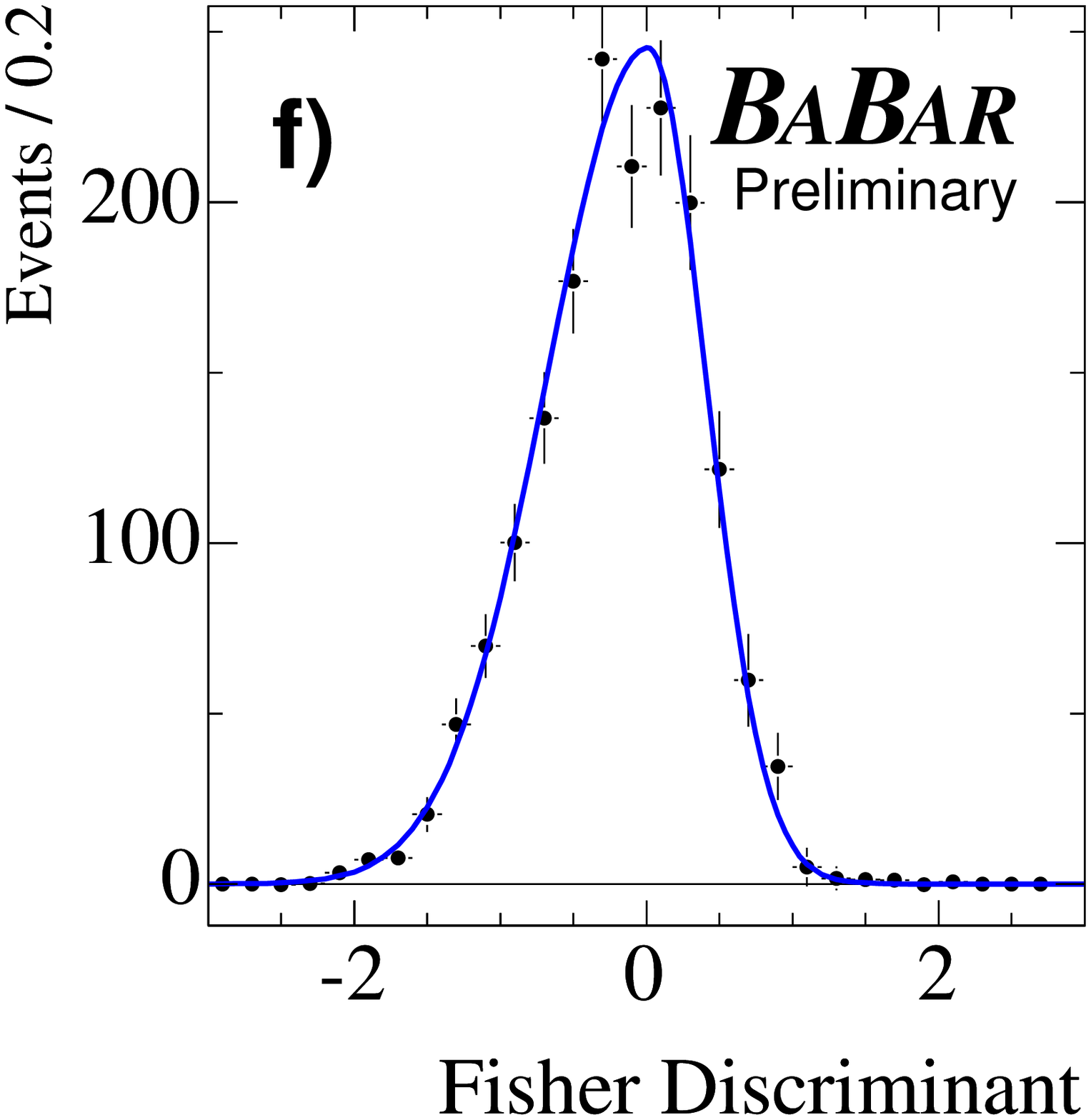}
\caption{Distributions (points with error bars) of $\mes$, $\de$, and ${\cal F}$ for signal 
$\pip\pim$ (a,b,c) and $\Kp\pim$ (d,e,f) decays using the weighting technique described in 
Ref.~\cite{sPlots}.  Solid curves represent the corresponding PDFs used in the fit.  The
distribution of $\de$ for $\Kp\pim$ events is shifted due to the assignment of the pion mass
for all tracks.}
\label{fig:signal}
\end{center}
\end{figure}

\begin{figure}[!tbp]
\begin{center}
\includegraphics[width=0.32\linewidth]{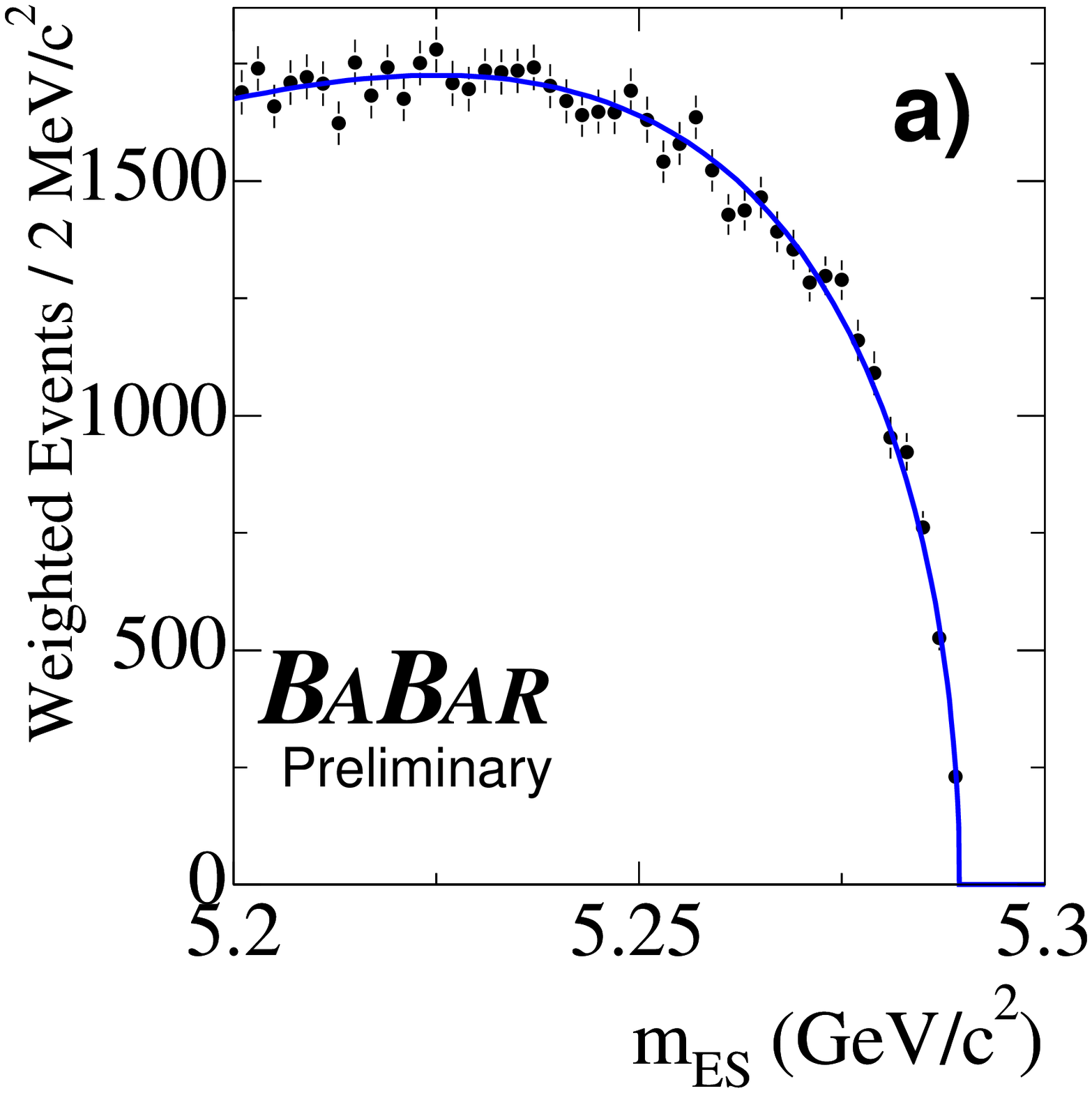}
\includegraphics[width=0.32\linewidth]{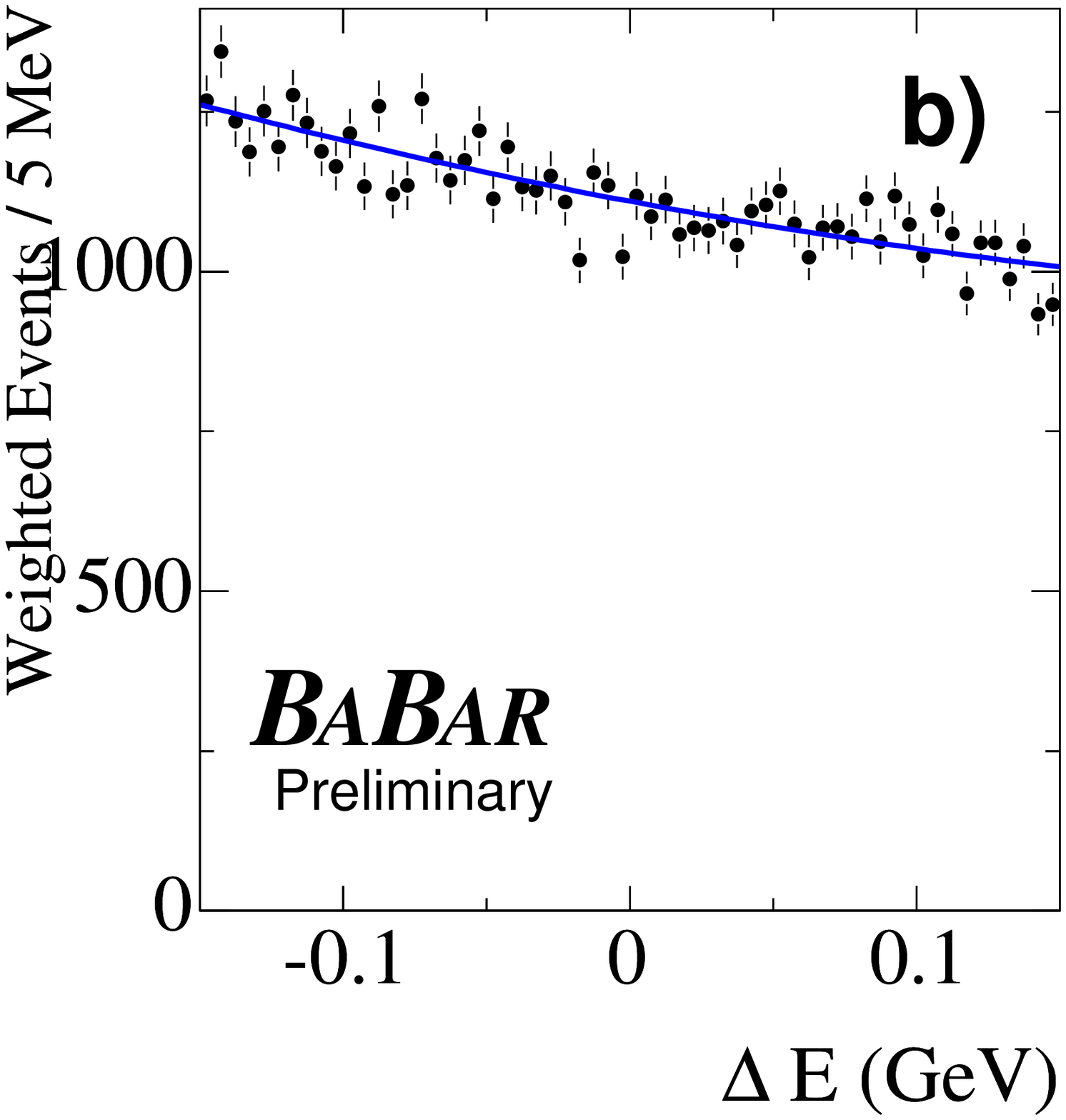}
\includegraphics[width=0.32\linewidth]{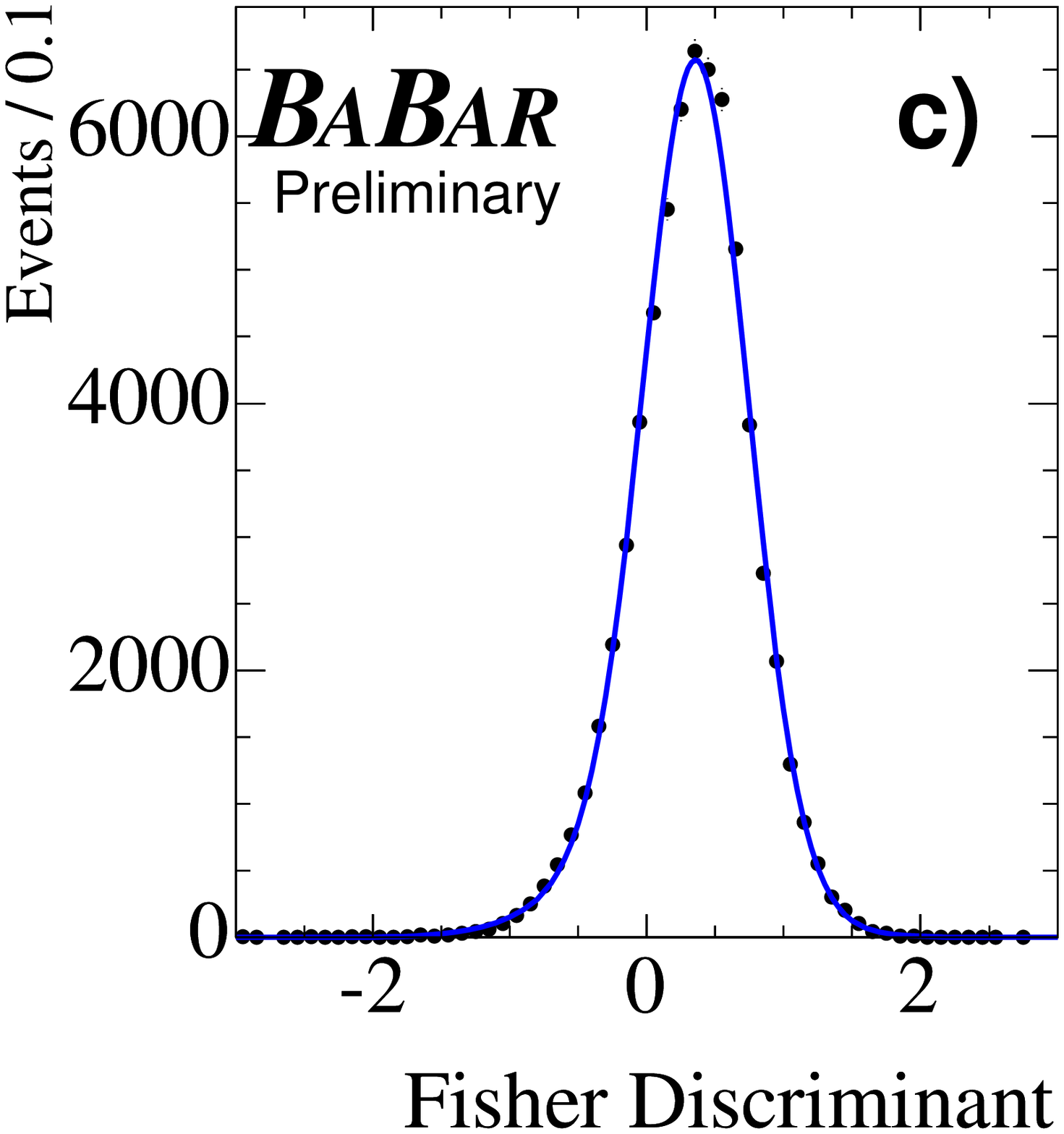}
\caption{Distributions of a) $\mes$, b) $\de$, and c) ${\cal F}$ for $q\bar{q}$ background events 
(points with error bars) using the weighting technique described in Ref.~\cite{sPlots}.  Solid 
curves represent the corresponding PDFs used in the fit.}
\label{fig:bkg}
\end{center}
\end{figure}

\begin{figure}[!tbp]
\begin{center}
\includegraphics[width=0.32\linewidth]{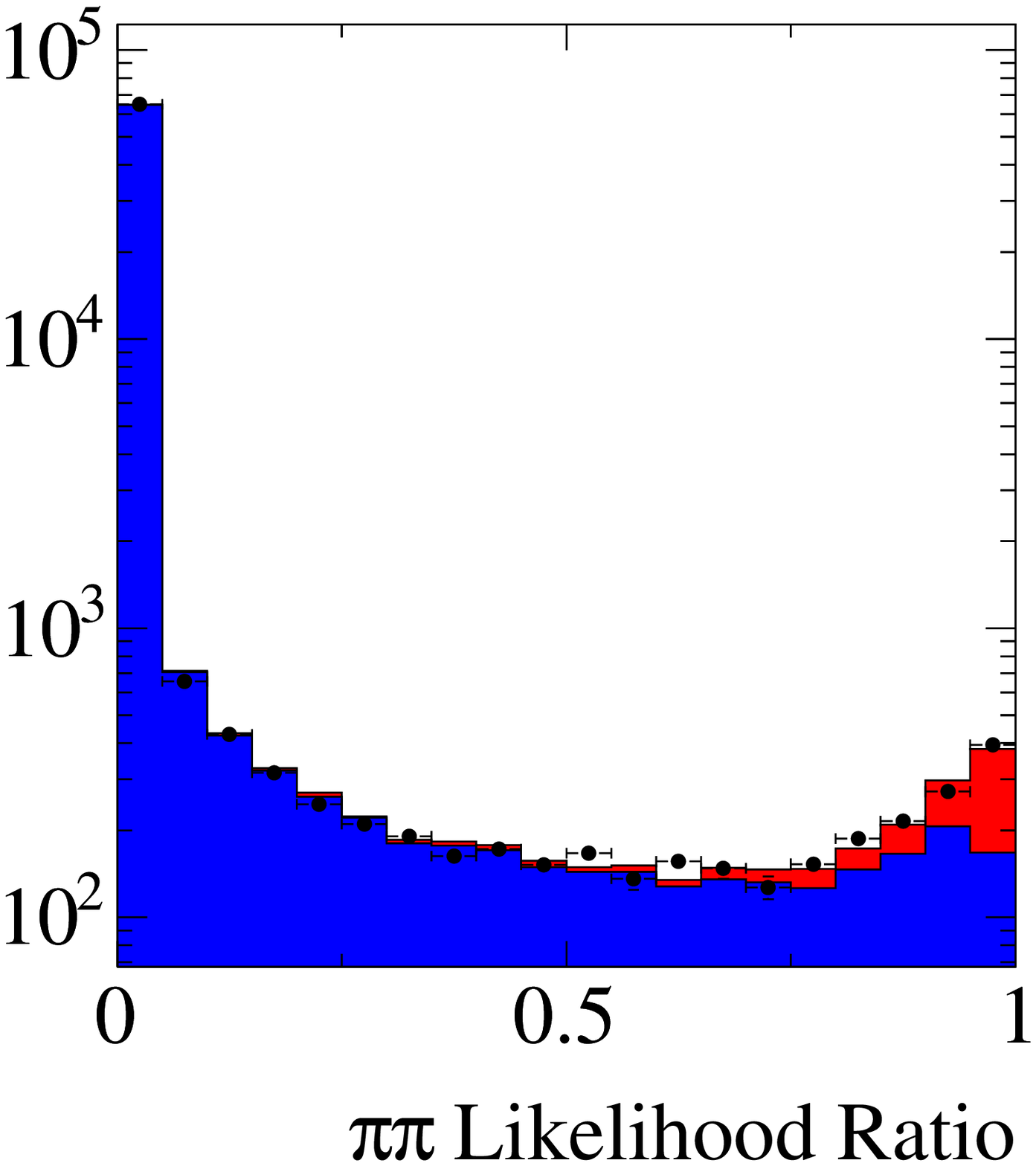}
\includegraphics[width=0.32\linewidth]{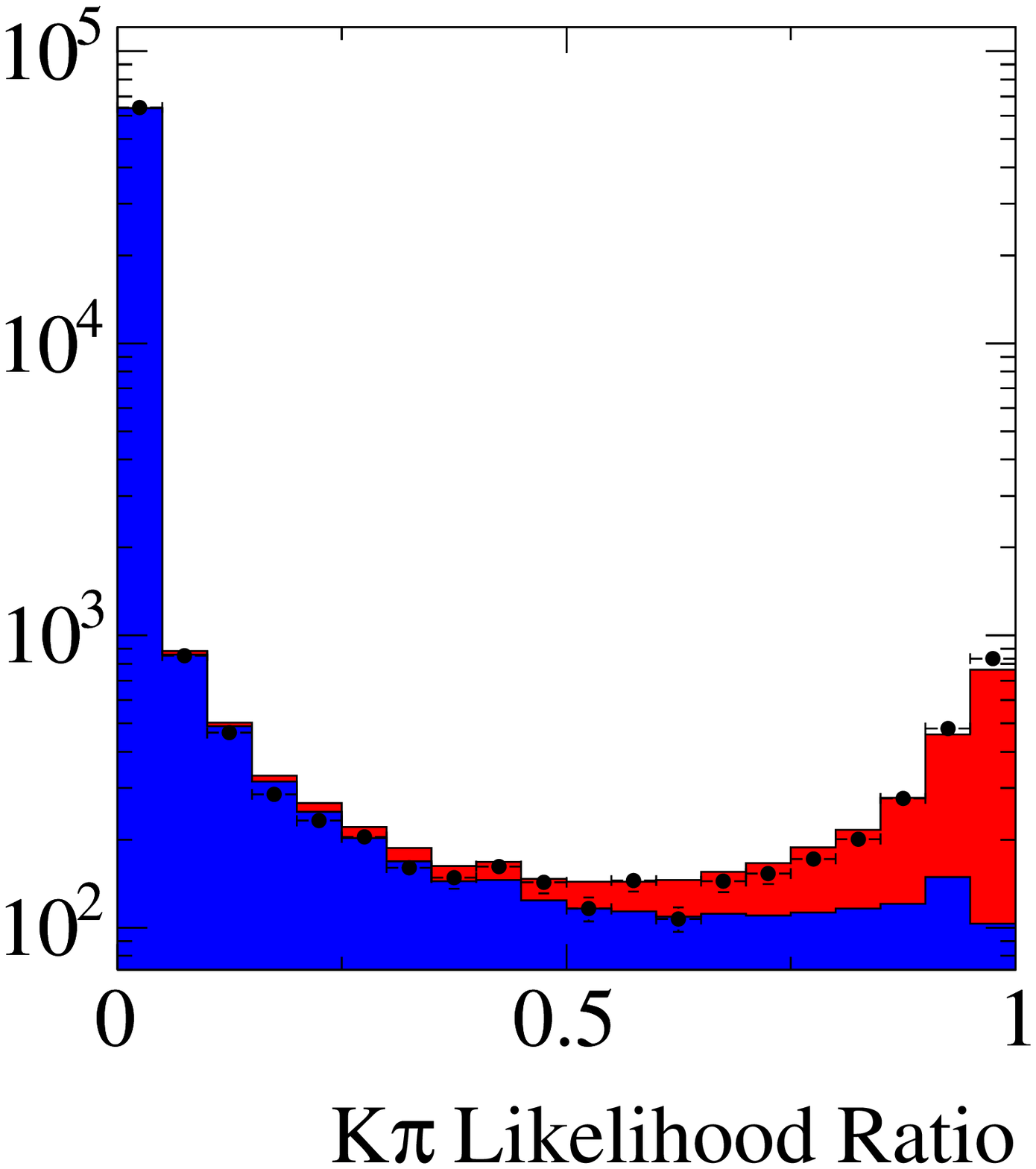}
\includegraphics[width=0.32\linewidth]{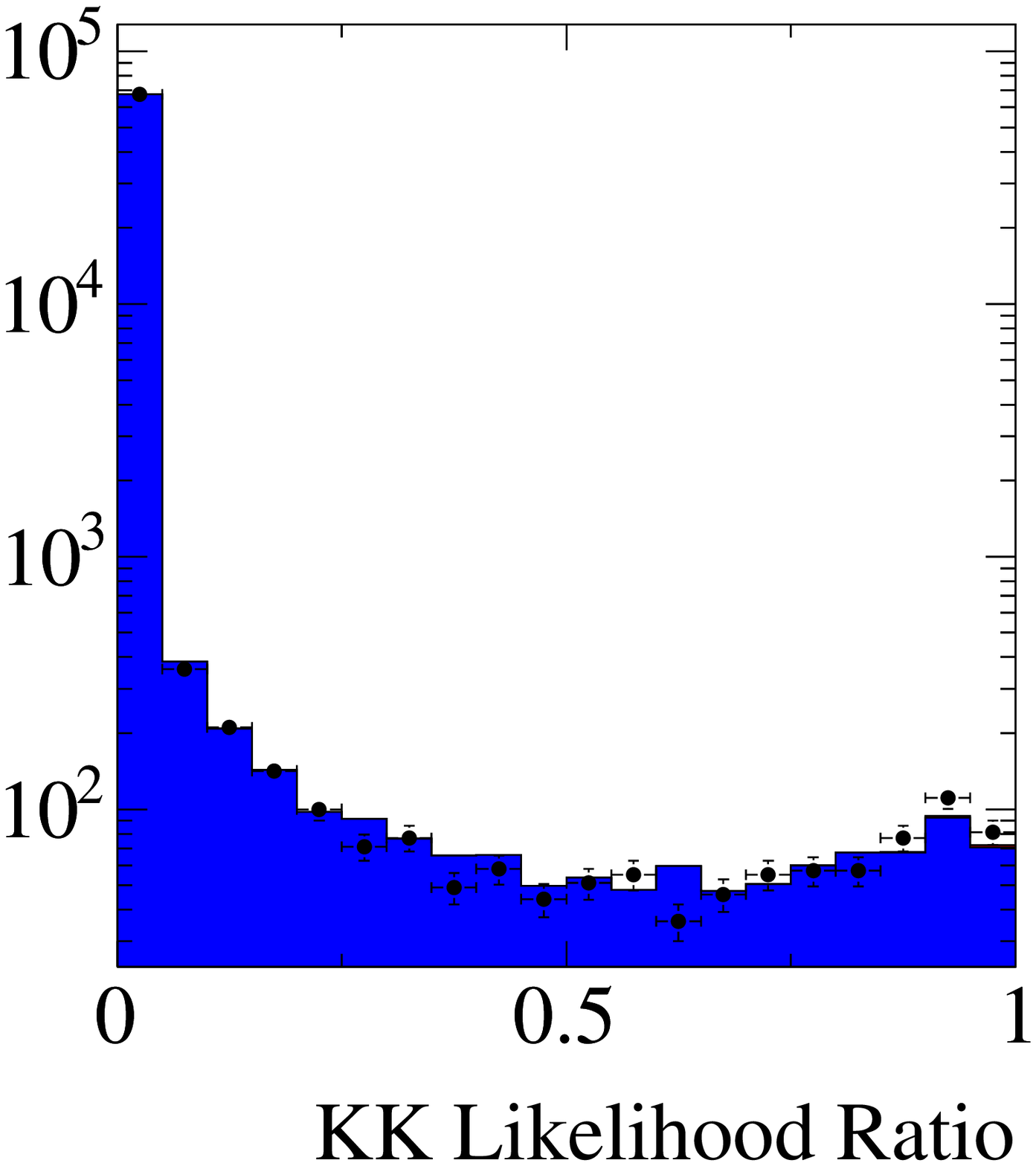}
\caption{The likelihood ratio ${\cal L}_S/\sum{{\cal L}_i}$, where ${\cal L}_S$ is the
likelihood for each event to be a signal $\pi\pi$ (left), $K\pi$ (middle), or $KK$ (right) event.  
The points with error bars show the distribution obtained on the fitted data sample, while the
histograms show the distributions obtained by generating signal (red) and background (blue) events
directly from the PDFs.}
\label{fig:lr}
\end{center}
\end{figure}

As a crosscheck, in Figs.~\ref{fig:signal} and~\ref{fig:bkg} we compare the PDF shapes
(solid curves) to the data using the event-weighting technique described in 
Ref.~\cite{sPlots}.  For each plot, we perform a fit excluding the variable being plotted and
use the fitted yields and covariance matrix to determine a weight that each event is either
signal or background.  The distribution is normalized to the yield for the given component and
can be compared directly to the assumed PDF shape.  For $\mes$, $\de$, and ${\cal F}$, we find 
excellent agreement for signal $\pi\pi$ and $K\pi$ events (Fig.~\ref{fig:signal}), as well as 
the sum of all channels for background events (Fig.~\ref{fig:bkg}).  We have verified separately 
that the background PDF shapes agree for all three channels.  Figure~\ref{fig:lr} shows the
likelihood ratio ${\cal L}_S/\sum{{\cal L}_i}$ for all $69264$ events in the fitted sample,
where ${\cal L}_S$ is the likelihood for a given signal hypothesis, and the summation in the
denominator is over all signal and background components in the fit.  We find satisfactory 
agreement between data (points with error bars) and the distributions obtained by directly 
generating events from the PDFs (histograms).

\section{SYSTEMATIC STUDIES}
Systematic uncertainties on the branching fractions arise from uncertainties on the selection
efficiency, signal yield, and number of $\BB$ events in the sample.  Uncertainty on the 
efficiency is dominated by track reconstruction ($1.6\%$) and the effect of FSR ($1.3\%$), which
is taken to be the difference between the efficiency as determined in the PHOTOS simulation and
the QED calculation (Table~\ref{tab:eff}).
Uncertainty on the fitted signal yields is dominated by the shape of the signal PDF for
${\cal F}$ ($2.9\%$ for $\pi\pi$, $1.5\%$ for $K\pi$) and potential bias 
($2.2\%$ for $\pi\pi$, $0.9\%$ for $K\pi$) in the fitting technique determined from large samples 
of Monte Carlo signal events and a large ensemble of pseudo-experiments generated from the PDF 
shapes.  Uncertainties due to imperfect knowledge of the PDF shapes for $\mes$, $\de$, and
$\theta_c$ are all less than $1\%$.  Table~\ref{tab:sysbr} summarizes the total uncertainty on 
the branching fractions, which is calculated as the sum in quadrature of the individual 
uncertainties.

\begin{table}[!tbp]
\caption{Summary of relative systematic uncertainties on yields, efficiencies, and 
number of $\BB$ pairs.  For the $\Kp\Km$ yield we show the absolute uncertainty.  
The total uncertainties for $\pip\pim$ and $\Kp\pim$ are calculated as the sum in 
quadrature of the individual contributions.}
\begin{center}
\begin{tabular}{cccc}
\hline\hline
Source           	& $\pip\pim$ & $\Kp\pim$ & $\Kp\Km$\\
\hline
yields                  &$ 3.8\% $&$ 1.8\% $&$ 6.8 $\\
efficiency              &$ 2.6\% $&$ 2.5\% $&$ 2.0\% $\\
$N_{\BB}$               &$ 1.1\% $&$ 1.1\% $&$ 1.1\% $\\
\hline
Total                   &$ 4.7\% $&$ 3.3\% $& n/a \\
\hline\hline
\end{tabular}
\label{tab:sysbr}
\end{center}
\end{table}

\section{RESULTS and SUMMARY}
\label{sec:Physics}
Table~\ref{tab:brfinalresults} summarizes the preliminary results for the charge-averaged 
branching fractions.  For comparison, we use the efficiencies and signal yields determined under 
the assumption of no FSR and find $\BR(\Bz\to\pip\pim) = 5.1\times 10^{-6}$ and 
$\BR(\Bz\to\Kp\pim)= 18.1\times 10^{-6}$, which are consistent with our previously published 
results~\cite{BaBarsin2alpha2002}.  Taking into account FSR effects leads to an increase of the 
branching fractions by approximately $8\%$ and $6\%$ for $\pi\pi$ and $K\pi$, respectively.
The upper limit on the signal yield for $KK$ is given by the value of 
$N_0$ for which 
$\int_0^{N_0} {\cal L}_{\rm max}\,dN/\int_0^\infty {\cal L}_{\rm max}\,dN = 0.90$, 
corresponding to a one-sided $90\%$ confidence interval.  Here, ${\cal L}_{\rm max}$ is the 
likelihood as a function of $N$, maximized with respect to the remaining fit parameters.  
We find $N_0 = 25.9$, and the branching fraction is calculated by increasing the signal 
yield upper limit and reducing the efficiency by their respective total errors 
(Table~\ref{tab:sysbr}).  For the purpose of combining with measurements by other experiments, we
have also evaluated the central value for the branching fraction and find
$\BR(\Bz\to\Kp\Km) = (4\pm 15\pm 8)\times 10^{-8}$.

\begin{table}
\caption{Summary of branching fraction results in a sample of $(226.6\pm
1.2)\times 10^{6}$ $\BB$ pairs.  
We show signal yields $N_S$, total detection efficiencies ($\epsilon$) and branching fractions
$\BR$ in units of $10^{-6}$.  The errors are statistical and systematic, respectively, and
the upper limit on $\Bz\to\Kp\Km$ corresponds to the $90\%$ confidence level.}
\begin{center}
\begin{tabular}{lcccc} 
\hline\hline
   Mode    &  $N_S$                  & $\epsilon\,(\%)$ & \BR($10^{-6}$) \\ 
\hline
$\pip\pim$ &  $491\pm 35\pm 11$      & $39.4\pm 0.2\pm 0.9$  & $5.5\pm 0.4\pm 0.3$ \\
$\Kp\pim$  &  $1674\pm 53\pm 15$     & $38.4\pm 0.2\pm 0.8$  & $19.2\pm 0.6\pm 0.6$ \\
$\Kp \Km$  &  $3.0\pm 13.1\pm 6.8\,(<25.9)$ & $37.6\pm 0.3\pm 0.8$  & $<0.40$ ($90\%$ C.L.) \\
\hline\hline
\end{tabular}
\end{center}
\label{tab:brfinalresults}
\end{table}

In summary, we have presented preliminary updated measurements of charge-averaged branching
fractions for the decays $\Bz\to\pip\pim$ and $\Bz\to\Kp\pim$, where FSR effects have been
taken into account.  We find a value of $\akpi$ consistent with the result in 
Ref.~\cite{babarAkpi}, and branching fractions $6$-$8\%$ higher due to the
effect of FSR on the efficiency and signal-yield determination.  This difference should be
taken into account when comparing with previous measurements of these quantities 
(Table~\ref{tab:oldresults}) that do not include these effects.  Our results are consistent
with current theoretical estimates using various techniques~\cite{thy}.  We find no 
evidence for the decay $\Bz\to\Kp\Km$ and set an upper limit of $4.0\times 10^{-7}$ at the 
$90\%$ confidence level.

\section{ACKNOWLEDGMENTS}
\label{sec:Acknowledgments}

% Standard acknowledgments paragraph; must always be included.
We are grateful for the 
extraordinary contributions of our \pep2\ colleagues in
achieving the excellent luminosity and machine conditions
that have made this work possible.
The success of this project also relies critically on the 
expertise and dedication of the computing organizations that 
support \babar.
The collaborating institutions wish to thank 
SLAC for its support and the kind hospitality extended to them. 
This work is supported by the
US Department of Energy
and National Science Foundation, the
Natural Sciences and Engineering Research Council (Canada),
Institute of High Energy Physics (China), the
Commissariat \`a l'Energie Atomique and
Institut National de Physique Nucl\'eaire et de Physique des Particules
(France), the
Bundesministerium f\"ur Bildung und Forschung and
Deutsche Forschungsgemeinschaft
(Germany), the
Istituto Nazionale di Fisica Nucleare (Italy),
the Foundation for Fundamental Research on Matter (The Netherlands),
the Research Council of Norway, the
Ministry of Science and Technology of the Russian Federation, and the
Particle Physics and Astronomy Research Council (United Kingdom). 
Individuals have received support from 
CONACyT (Mexico),
the A. P. Sloan Foundation, 
the Research Corporation,
and the Alexander von Humboldt Foundation.


\begin{thebibliography}{99}
\bibitem{cc}
The use of charge conjugate modes is implied throughout this paper unless otherwise
stated.
\bibitem{ref:mesdedef}
The variables $\de$ and $\mes$ are defined in Ref.~\cite{babarAkpi}.
\end{thebibliography}
\end{document}